# Broadband Terahertz Wave Emission from Liquid Metal

Yuqi Cao,[1, 2] Yiwen E,[1] Pingjie Huang,[2] and X.-C. Zhang [1, a)]

[1] *The Institute of Optics, University of Rochester, Rochester, NY 14627, USA*

[2] *College of Control Science and Engineering, Zhejiang University, Hangzhou 310027, China*

[a)] *Corresponding author: xi-cheng.zhang@rochester.edu*

Metals have been studied as terahertz sources for decades. Recent research has shown the potential of metals in generating extremely high THz pulse energy excited by intense laser pulses. To avoid the metal surface debris caused by laser pulses, here, we report the results of the broadband terahertz wave emission from a flowing liquid metal line excited by sub-picosecond laser pulses. The THz signal emitted from the liquid gallium line shows stronger field with broader bandwidth comparing with the signal from water under the identical optical excitation conditions. Our preliminary study suggests that the liquid metals have the potential to serve as efficient and powerful THz sources for the intense lasers with a high repetition rate.

Over the last decade, several new terahertz (THz) frontiers such as condensed-matter dynamics,[1] strong THz field-matter interaction,[2-4] imaging applications,[5] and nonlinear THz spectroscopy,[6-8] have attracted considerable interest. To take these opportunities requires the development of powerful and efficient THz sources. One way is to use laser-induced plasma excited by ultra-intense laser pulse on solid or gas targets.[2, 9-10] Gopal et al. reported the 0.7 mJ THz radiation generation from the rear surface of a 5 µm thick stretched titanium metal foil with an intense femtosecond laser pulse, wherein the sheath acceleration dominates the ion acceleration process.[9] Also, an electric field of over 8 MV/cm of THz waves excited by two-color air-plasma was obtained with the help of combining a thin dual-wavelength half-waveplate and a Brewster-angled silicon window.[2]

Metallic targets have been considered as great THz radiation sources because of their relative lower ionization thresholds and higher molecule densities compared with gas, which allows THz wave generation using lower pump energy.[11] In addition, there is no need to consider the phonon absorption comparing to solid crystals.[12] Early studies on THz wave generation from metallic targets have been widely reported.[4, 9, 13-17] Specifically, an intense and coherent THz wave emission with a pulse energy is as high as 55 mJ from a copper foil target irradiated by a high-intensity picosecond laser pulse has been released recently.[4] However, the debris created by the intense laser pulses on the metal surfaces limits its further application. Thus, better targets are highly required to match the laser with high repetition rate. Liquid metal is taken into consideration because they can provide a pristine, unperturbed smooth surface repeatable for continuous laser shot.

In this letter, we report the use of a flowing liquid gallium (LG) line as a THz radiation emitter. The THz signal generated from LG by single-color optical excitation is collected in a transmission geometry. THz waves emission from air, liquid water and liquid metal with identical excitation conditions are measured. THz signals from liquids (LG and water) are stronger than that from air. Also, a 1.7 times enhancement of THz peak field is obtained from the LG line compared with that from water. Similar polarity flips of THz electric field according to the relative position between the incident laser and target liquid line are observed for both cases of LG and water. In addition, it is measured that a longer optical pulse duration works better for LG, indicating that the ionization process is likely attributed to the cascade ionization. The ionized electrons are accelerated by the laser ponderomotive force and THz wave is radiated when they pass through the metal-air interface.

For the targets, we consider the metals with relative low melting points and convenience to obtain. **Table 1** lists the melting point, surface tension, density, viscosity, and ionization energy of the candidates at certain temperatures. The most commonly used liquid metal is the mercury because of the lowest melting point among all metals, however, it is not considered in our experiments due to the toxicity concern. Among all metals in this table, liquid gallium is a promising candidate for a THz source since the melting point of gallium (Ga) is just above the room temperature (30 °C). Also, compared with other candidates listed in the table, Ga owns better chemical stability and physical safety, which make it an ideal liquid metal target and widely used in x-ray generation applications as well.[18,19]

The surface tension of LG at 35 °C is 0.708 N/m, which is more than 10 times higher than that of water (0.0690 N/m) at the same temperature.[23] The high surface tension helps to form a smoother surface of a flowing liquid line. Furthermore, the density of LG (6.109 g/cm$^3$ at 35 °C) is larger than that of solid gallium (5.91 g/cm$^3$).[21] A solid Ga ingot with 99.998% purity was melted into and maintained at the liquid phase at 35 °C during all measurements.

The optical setup of THz wave generation from liquid metal employs a Ti:sapphire amplified laser with 800 nm center wavelength and 1 kHz repetition rate. The p-polarized laser pulse with 0.4 mJ energy and 370 fs pulse duration is focused by a 50 mm focal length lens at the liquid metal line (LML). The laser focal size on the LML and the Rayleigh length of the laser beam are 5 µm and 51 µm, respectively. The schematic diagram of the setup is shown in **Fig. 1(a)**. The movement of the liquid line near the focus along x and z direction is finely controlled by a two-dimensional translation stage, respectively. The forward propagating THz signal is detected by a standard electro-optical sampling using a 3 mm thick <110> cut ZnTe crystal. More details of the experimental setup can be found in an earlier work.[24] A commercial peristaltic pump is used to create a LG line flowing at 3.8 m/s steadily with a diameter of 210 µm as shown in **Fig. 1(b)**. As the flow rate of the liquid metal is greater than 1m/s, the liquid target at the focal point is replenished and restored after each laser pulse under the 1 kHz laser repetition rate.

The temporal THz waveforms from air, water, and LG are recorded under the same optical excitation conditions, as shown in **Fig. 2(a)**. Note the signal from air plasma is enlarged by 10 times in the plot. When optical pulse duration is varied from 35 fs to 420 fs, we find that the optimal pulse durations for LG and water are both 370 fs. All the THz signals in **Fig. 2(a)** are measured with the identical optical pulse duration 370 fs and pulse energy. The diameter of the liquid water line and the LG line are produced by a same needle with an inner diameter of 210 µm, and the laser focal position on the two liquid lines are optimized respectively to obtaining the maximum THz field in the forward direction.

Remarkably, THz signal from the LG is much stronger than the signal from the air. Considering air-plasma which prefers to a shorter pulse duration, the THz signal from air excited by shorter pulse is also measured and the result is still one-order weaker than the LG signal. The THz electric field generated from ionized LG is 1.7 times stronger than that from water at 35 °C. This might be attributed to the higher density of ionized electrons from LG line. The corresponding comparison in the frequency domain is plotted in **Fig. 2(b)**. It is noteworthy that the THz signal from LG has a broader bandwidth than that from water and a stronger spectral amplitude in high frequency. The spectrum of LG contains more high frequency components than water spectrum because of the high absorption of THz radiation by liquid water at higher frequency. The frequency bandwidth of THz signal emitted from the LG l obtained in our setup is limited by the detector phonon bandwidth of crystal ZnTe.

To further study the THz field strength dependence on the focal location on LG line, the LG line is scanned along x-axis. Note the z position of the 2-dimensional stage is optimized for the THz field previously. The normalized THz strength of both water and LG lines as a function of x position are plotted in **Fig. 3**. When x equals to 0, the optical pump pulse is focused at the center of LG line. The curves show that both THz signals generated from water and LG possess two peaks, but with reversed polarity. This is caused by the flipped dipole for the opposite incident angle.[25] The dashed black and red lines in the figure illustrate the position of the peak THz fields of water and LG, and the separation between the two maxima of the LG signal is larger than the case of water. This means that the positions where the peak THz fields are emitted from the LG line are closer to the edge of the liquid line. Note that the diameter of the liquid line and the laser focused spot size are 210 µm and 5 µm, respectively. Thus, in the case of LG, only part of the optical pulse is focused at the edge of the liquid line when the peak THz field is generated, which might be explained by the low penetration depth of LG for both optical and THz wavelengths.

The mechanism of THz wave from LG line might be attributed to the coherent transition radiation when the laser-induced electrons pass through the metal-air interface.[26, 27] The skin depth of LG at both optical (800 nm) and THz (∼300 µm) wavelengths are estimated to be around 7.7 nm and 60 nm respectively, which is much less than the diameter of LG line (210 µm). Thus, ionization starts from the surface at LG. Since a longer pulse duration works better for THz wave generation from LG, like water, enough seed electrons are ionized through multiphoton ionization at the front part of the optical pulse. Then cascade ionization occurs.[25, 28] The mean-free-time of electrons in ionized LG is estimated to be less than 1 fs, which is much shorter than the laser pulse duration. This indicates that the collisions of electrons continuously occur for a long period of time and leads to an exponential increase in the number of electrons.[28] Furthermore, another possible reason for the THz radiation emission enhancement from LG than water is the large number of electrons ionized from LG due to the higher molecular density of gallium. After the ionization process, the electrons are accelerated by the ponderomotive force as a result of the non-uniform density gradient distribution of plasma, and THz wave is emitted when the electrons cross the boundary of LG and air.

In summary, we report the preliminary experimental data to reveal the possibility of strong THz emission from the LG line excited by sub-picosecond laser pulses. Although the flip of THz waveforms from LG observed by scanning along x-axis shows similar characteristics with water signal, the mechanism of THz waves generation from liquid metals likely differ from the generation process of liquid water. Further experiments still need to be conducted. THz wave emission pattern, especially sideway and backward direction, is an important experiment to understand the THz radiation mechanism. In addition, the THz wave emission from different liquid metal targets will be investigated to acquire a strong THz source. One of other good liquid metal candidates is Galinstan, which is a eutectic alloy with its composition of gallium, indium, and tin. This eutectic alloy, with its melting between −19 °C to +11 °C dependent on the ratio of compositions, is at liquid state at room temperature. Our results suggest that liquid metal targets have the potential to be strong THz sources for intense lasers with a high repetition rate and may provide a new perspective to study the physics of laser-metal interactions.

**Acknowledgments**

The research at the University of Rochester is sponsored by the Army Research Office under Grant No. W911NF-17-1-0428, Air Force Office of Scientific Research under Grant No. FA9550-18-1-0357, and National Science Foundation under Grant No. ECCS-1916068. Cao acknowledges helpful discussions with Kareem J. Garriga Francis on intense laser and liquid metal interaction process. Zhang thanks Alexander Shkurinov's suggestion and discussion.

**Data availability**

The data that support the findings of this study are available from the corresponding author upon reasonable request.

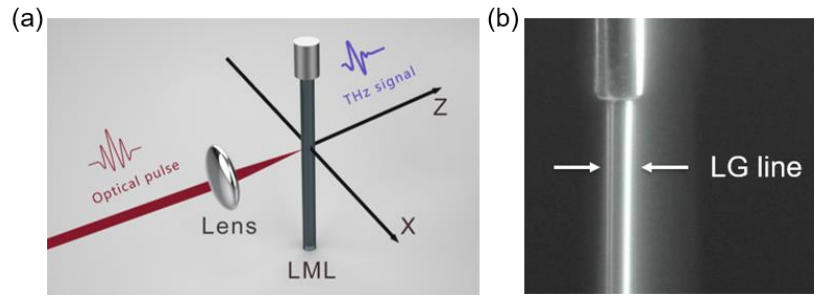

Fig. 1. (a) Schematic of the experimental setup. A two-dimensional translation stage is used to control the position of x-axis and z-axis of the LML. Optical beam propagates along the z-axis and THz emission is measured in the forward direction. (b) Photo of a flowing LG line in ambient air. The diameter of the LG line is 210 µm. The high surface tension of LG makes an excellent surface quality.

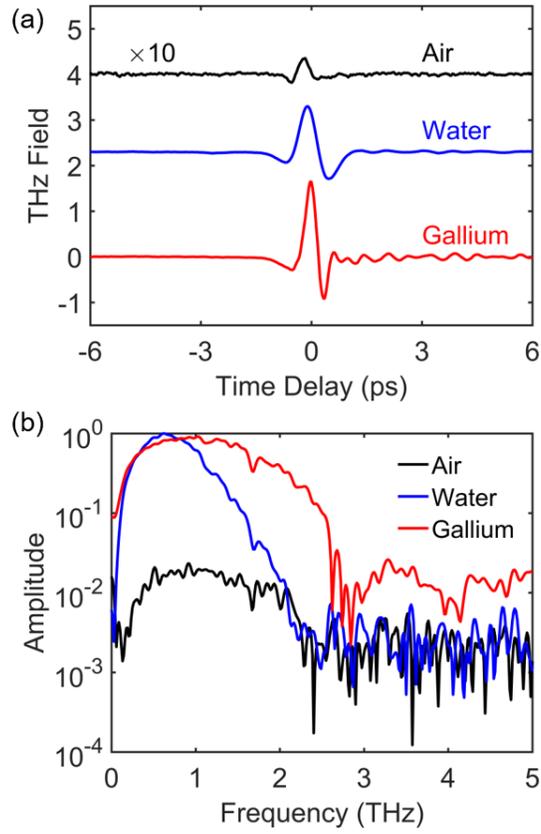

Fig. 2. Comparison of THz waves generated from air, and a 210µm diameter lines of water and gallium with single-color excitation scheme. (a) Comparison of THz field strengths in air plasma, water and liquid gallium. The THz signal from single color air plasma is enlarged by 10 times. (b) Corresponding comparison in the frequency domain.

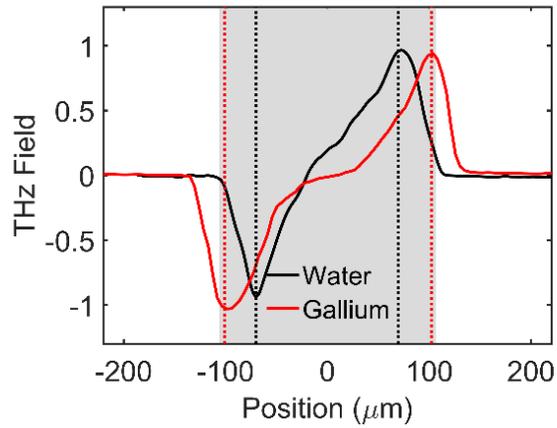

Fig. 3. Normalized THz field strength as a function of the x-axis position from liquid water and LG lines. The dashed black and red lines represent the x-axis position of the maximum THz fields for water and LG, respectively. The width of the grey area stands for the diameter of the LG line. The positions for obtaining the maximum THz field strengths for water and LG are 70 µm and 101 µm, respectively.

Table 1. Physical properties for selected metals.[20-22] The surface tension, density and viscosity of each candidates are at the temperature listed in the first column.

| Metal Targets | Melting point (°C) | Surface tension (N/m) | Density (g/cm3) | Viscosity (Pa·s) | Ionization energy (eV) |
|---|---|---|---|---|---|
| Mercury (Hg) (at 20°C) | −38.8 | 0.487 | 13.534 | 0.0015 | 10.44 |
| Cesium (Cs) (at 60°C) | 28.5 | 0.675 | 1.843 | 0.0058 | 3.90 |
| Gallium (Ga) (at 35°C) | 29.8 | 0.708 | 6.109 | 0.0020 | 5.98 |
| Rubidium (Rb) (at 100°C) | 39.3 | 0.854 | 1.460 | 0.0048 | 4.18 |
| Phosphorus (P) (at 50°C) | 44.0 | 0.698 | 1.740 | 0.0016 | 10.64 |
| Indium (In) (at 170°C) | 156.6 | 0.558 | 7.020 | 0.0019 | 6.08 |